# Developing Assessment Methods for Evaluating Learning Experience

**Maneesha**[0000-0002-4036-5982]

[1]Department of General Science, BITS Pilani Dubai Campus, Dubai, UAE

maneesha@dubai.bits-pilani.ac.in

**ABSTRACT**

This research aims to investigate the gender-based learning experiences of engineering students enrolled in the Probability and Statistics course, focusing on the four different assessment methods employed namely direct conceptual learning (DCL), symposium, applied deployment and collaborative learning. The study encompasses 299 engineering students, comprising 90 females and 209 males. Multivariate Analysis of Variance (MANOVA), is used to gain deeper insights into the complex interplay between assessment methods and their influence on student learning. The results of the statistical analysis reveal that there are significant differences in the learning outcomes between female and male engineering students in the assessment methods of direct conceptual learning, symposium, and applied deployment. The findings suggest that there is no significant difference in the learning outcomes between female and male engineering students in the collaborative learning assessment method. The graphical representation visually confirms the significant differences in direct conceptual learning, symposium, and applied deployment, while illustrating no significant difference in collaborative learning between female and male engineering students.

## 1. INTRODUCTION

The education sector has been working on improving assessment practices in higher education across the world, considering the fact that appropriate assessment methods play a vital role in achieving the goal of education in any country. Studies on classroom assessment are an essential aspect of effective teaching and learning [1], [2]. Assessment is one of the core elements important to evaluate the learning performance of students. Ibarra-Sáiz et al. [3] discussed a causal relationship model including key variables such as participation, self-regulation, learning transfer, strategic learning, feedback, and empowerment. The results highlight how assessment practices in higher education can be enhanced through improvements in the design of assessment. Designing the assessment process involves making decisions to determine its purposes, what the learning outcomes will be, its context, how feedback will be organized, and, of course, what assessment tasks will be undertaken [4]. To enhance the learning experience, Information and Communication Technology (ICT) should be used in teaching and learning Business Education in the University [5].

For decades, several educators have viewed assessment as a means for measuring final learning outcomes and this is mainly actualized through what is known as summative assessment [6]. Further, the researchers explored how teachers may develop productive relationships between the formative and summative functions of classroom assessment, so that their judgments may inform the formal external assessment of students, thus increasing the validity of those assessments [7]. However, educators have now started widening their scope of assessment to cover not only students' learning outcomes at the end of a specific period to decide who passes or fails, but also to enhance learning by modifying classroom instruction [8]. Educators have further pondered that with the advent of the digital age, traditional didactic teaching and online learning have been modified and gradually replaced by "Blended Learning." [9] Blended learning has a positive impact on increasing students' academic achievement [10].

Engineering students must possess the subject knowledge and skills to become engineering professionals. In this viewpoint, assessment plays a vital role in assisting students in knowing about their knowledge and skills in engineering courses [11]. Burtner [12] argued that engineering students require certain skills (such as analytical skills, problem-solving skills, communication skills, teamwork skills, etc.) that would not be assessed adequately by traditional assessment practices. There is a need for the development and implementation of more effective assessment practices in engineering education settings.

The nature and scope of assessment in engineering education are influenced by several specific traits that distinguish engineering studies from other disciplines. These traits highlight the unique challenges and objectives of engineering education. Engineering students are expected to acquire theoretical and technical knowledge in their respective engineering disciplines. They are expected to possess high-quality oral and written communication skills to convey their knowledge [13]. According to Cruz [14], engineering students are supposed to attain the ability to work as competent team members in an interdisciplinary collaborative environment. Eventually, students are expected to get hold of engineering practices to evolve as engineering professionals [15], [16]. Assessment and feedback practices can continuously monitor and improve engineering students' progress in all these requirements.

Researchers have interpreted engineering learning in several ways: *Design practice* allows students to apply theoretical knowledge to real-world scenarios and develop practical problem-solving skills and a hands-on approach helps students develop a deeper understanding of engineering principles and fosters creativity [17], [18]. *Interactive reflection*, in which engineering students interact with people and objects as a means of solving design problems [19], [20] via frequent and timely reflections and discussions that are constantly summarized to form deeper understandings. *Knowledge integration happens when* learners receive external information, integrate this with their internal knowledge, and transform both into problem-solving strategies [21], [22], [23]. *Circular iteration*, with students continuously trying out and updating their designs until they verify the overall design solution. These scholars concur that, in engineering learning, the repetition of practice is critical because it provides them with the opportunity to test and apply their modified design solutions, explain their experiments, and learn from them [18], [24], [25], [26].

Another aspect is the perspective of the educator, in reflecting on the challenges, but also the scope, to include social learning tools as part of higher education teaching-learning. It identifies the preferences of Generation Z learners in terms of social media tools, and the trends thereof [27]. Further, there is emerging evidence to suggest that interactive apps may be useful and accessible tools for supporting early academic development. This is more applicable in early learning [28]. Teachers, instructors, and instructional designers may explore the free digital tools to further support student learning in their discipline. The notion of video games as a learning tool was also introduced [29], [30], [31].

However, a gap exists in the literature regarding the gender-based learning experiences of engineering students [32], [33]. The limited focus of researchers on the differences between the learning outcomes of male and female engineering students with respect to different assessment methods indicates the need to carry out extensive research on the topic. The present research attempts to fill this gap and come up with strong conclusions regarding the importance of assessment methods for evaluating learning experiences among male and female engineering students.

This study examines gender-based learning among engineering students. The assessment methods are developed incorporating design practice, interactive reflection, knowledge integration, circular iteration, blended learning, and the use of digital tools. These facilitate a learning experience where students utilize their prior knowledge and experiences to develop tangible models that tackle real-world engineering challenges. Through multiple iterations, students acquire new knowledge and enhance their problem-solving skills.

## 2. METHOD

### 2.1 Data

The data pertains to the performance scores of engineering students who were enrolled in the Probability and Statistics course. The performance scores are selected since these accurately point out the overall performance without bias. The four assessment methods were used to improve the student's learning experience. These are:

2.1.1. Direct conceptual learning: The aim was to assess the understanding of the concepts and the theory taught in the class and its application in real-world scenarios. This is to assess the design practice and blended learning experience. It was significant to determine the core aspects associated with direct conceptual learning.

2.1.2. Symposium: A real-life problem was selected. The discussion was initiated to come to a consensus as to which statistical method is best suited for analysis. The aim was that the learner get clarity on the application of statistical tools in different scenarios. Interactive reflection and digital tools have formed the basis of the assessment.

2.1.3. Applied deployment: It was an assignment for the student to individually solve a problem using any of the statistical software or programming language. The idea was to assess knowledge integration and the use of digital tools. The focus is on digital tools since these have become common among educators and students.

2.1.4 Collaborative learning: It was aimed to hone the team spirit amongst the students along with developing an interdisciplinary collaborative environment. An engineering problem was taken and analyzed by team members from different engineering disciplines. Circular iteration forms the basis of the assessment method.

The study involved a total of 299 engineering students, with 90 (approximately 30%) being females and 209 (approximately 70%) being males. The dataset has the scores of the students along with the gender specification. The names and other details of the students have not been disclosed thereby ensuring privacy protection and adhering to the ethical standards in research.

**2.2 Statistical Design**

The study employs a statistical design known as multivariate analysis of variance (MANOVA). The dependent variables under investigation are the scores obtained in direct conceptual learning, applied deployment, symposium, and collaborative learning. The independent variable being examined is the gender-based learning of engineering students. The primary hypothesis is:

$H_0$: There is no difference in the learning outcomes between male and female engineering students in the assessment methods of direct conceptual learning, applied deployment, symposium, and collaborative learning.
$H_1$: There is a significant difference in the learning outcomes between male and female engineering students in the assessment methods of direct conceptual learning, applied deployment, symposium, and collaborative learning.

**2.3 Assumptions** The following assumptions are met:

**2.3.1 Multivariate Normality** – This refers to the univariate normal distribution's generalization to two or more variables. Response variables are multivariate normally distributed within each group of the factor variable(s). Specifically, the response variables refer to dependent and independent variables.

**2.3.2 Independence** – Each observation is randomly sampled from the sample. Also, each observation is independently sampled from the population. The reason to ensure independent sampling is to ensure transparency and avoid bias in the research.

**2.3.3. Equal Variance** – A MANOVA assumes that the population covariance matrices of dependent variables direct conceptual learning, symposium, applied deployment and collaborative learning are equal between male and female engineering students. To check this assumption, we use Box's M test as shown in Table 1. This test is known to be quite strict, so we usually use a significance level of .001 to determine whether or not the population covariance matrices are equal.

**Table 1: Box's Test of Equality of Covariance Matrices**

| Box's M | 19.212 |
|---|---|
| F | 1.886 |
| df1 | 10 |
| df2 | 142979.769 |
| Sig. | .042 |

The p-value for Box's M test is 0.042 which is greater than .001, we can assume that this assumption is met.

**2.3.4. No Multivariate Outliers** – There are no extreme multivariate outliers. The most common way to check this assumption is to calculate the Mahalanobis distance for each observation, which represents the distance between two points in a multivariate space. If the corresponding p-value for a Mahalanobis distance of any observation is less than .001, the observation is considered to be an extreme outlier.

**3. RESULTS AND DISCUSSION**

**3.1 Descriptive Statistics**

Table 2 represents the descriptive statistics. In terms of the assessment methods of direct conceptual learning, symposium, applied deployment, and collaborative learning, the average scores of female engineering students surpass those of their male counterparts. Notably, the highest combined average score across all assessment methods is observed in collaborative learning, whereas the lowest average score is found in the symposium.

**Table 2: Descriptive Statistics**

|  | GENDER | MEAN | ST.DEV. | N |
|---|---|---|---|---|
| DCL | FEMALE | 61.1852 | 25.33660 | 90 |
|  | MALE | 48.3222 | 29.76039 | 209 |
|  | Total | 52.1940 | 29.06620 | 299 |
| SYMPOSIUM | FEMALE | 50.5554 | 30.42565 | 90 |
|  | MALE | 41.3237 | 31.44200 | 209 |
|  | Total | 44.1025 | 31.37665 | 299 |
| AD | FEMALE | 53.0190 | 24.48890 | 90 |
|  | MALE | 42.1293 | 28.60383 | 209 |
|  | Total | 45.4071 | 27.84276 | 299 |
| CL | FEMALE | 78.2962 | 20.49504 | 90 |
|  | MALE | 73.6522 | 23.63176 | 209 |
|  | Total | 75.0501 | 22.79914 | 299 |

**3.2 Multivariate Tests**

Wilk's Lambda is utilized to test the null hypothesis as shown in Table 3. The Significance value is the most important element in Wilk's Lambda test. The significance level of 0.011, being lower than the standard threshold of 0.05, indicates that there is indeed a significant difference in the performance scores of male and female students in the assessment methods employed.

**Table 3: Multivariate Tests**

| Effect |  | Value | F | Hypothesis df | Error df | Sig. |
|---|---|---|---|---|---|---|
| GENDER | Pillai's Trace | .043 | 3.343[b] | 4.000 | 294.000 | .011 |
|  | Wilks' Lambda | .957 | 3.343[b] | 4.000 | 294.000 | .011 |
|  | Hotelling's Trace | .045 | 3.343[b] | 4.000 | 294.000 | .011 |
|  | Roy's Largest Root | .045 | 3.343[b] | 4.000 | 294.000 | .011 |

**3.3. Levene's Test of Equality of Error Variances**

Levene's test is employed to examine the null hypothesis as represented in Table 4. It suggests that the error variance of the dependent variable is the same across groups. For symposium and collaborative learning, the significance levels are 0.547 and 0.268, respectively, both exceeding the 0.05 level. This indicates that the error variances are equal in these two groups. However, for direct conceptual learning and applied deployment, the significance levels are 0.012 and 0.025, respectively, which are less than the 0.05 level. Thus, we reject the null hypothesis, suggesting that the error variances are not equal across these two groups.

**Table 4: Levene's Test**

|  |  | Levene Statistic | df1 | df2 | Sig. |
|---|---|---|---|---|---|
| DCL | Based on Mean | 6.443 | 1 | 297 | .012 |
| SYMPOSIUM | Based on Mean | .363 | 1 | 297 | .547 |
| AD | Based on Mean | 5.074 | 1 | 297 | .025 |
| CL | Based on Mean | 1.230 | 1 | 297 | .268 |

### 3.4 Tests of Between-Subjects Effects

Table 5 represents the tests of Between-Subjects Effects. It represents the Significance F value for each dependent variable including direct conceptual learning (DCL), symposium, applied development (AD), and collaborative learning (CL). The focus is to examine the difference between female and male engineering students in terms of learning outcomes.

**Table 5: Between- Subjects Effects**

| Source | Dependent Variable | Type III Sum of Squares | df | Mean Square | F | Sig. |
|---|---|---|---|---|---|---|
| GENDER | DCL | 10408.870 | 1 | 10408.870 | 12.809 | .000 |
| | SYMPOSIUM | 5361.504 | 1 | 5361.504 | 5.529 | .019 |
| | AD | 7460.207 | 1 | 7460.207 | 9.911 | .002 |
| | CL | 1356.741 | 1 | 1356.741 | 2.624 | .106 |

#### 3.4.1 Direct Conceptual Learning:

The significance level for direct conceptual learning is 0.000 (Table 5), which is below the 0.05 level. This indicates a significant difference in the learning outcomes between female and male engineering students. The estimated margin levels are found to be 60 among females and close to 50 among males as shown in Fig 1. The graphical representation below confirms and supports the significant difference in the learning outcomes between males and females in the case of direct conceptual learning.

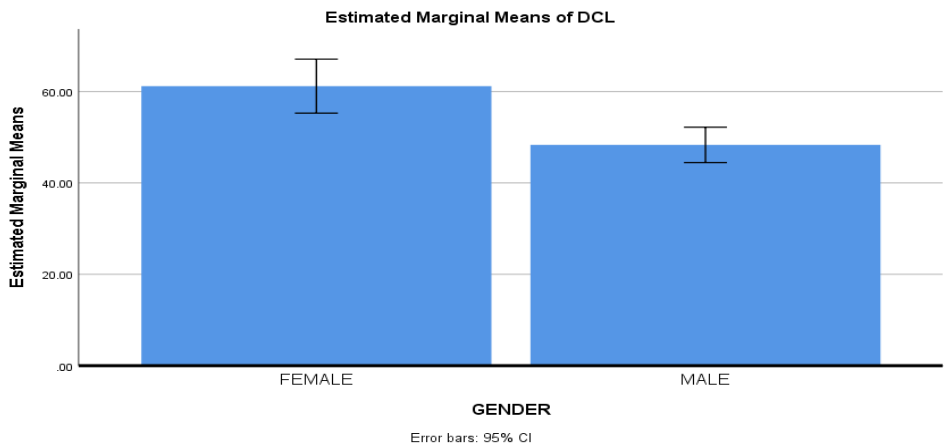

**Figure 1: Estimated Marginal Means of Direct Conceptual Learning**

#### 3.4.2 Symposium

The significance level for the symposium is 0.019 (Table 5), which is below the 0.05 level. This indicates a significant difference in the learning outcomes between female and male engineering students. The estimated margin levels are found to be 50 among females and 40 among males as shown in Fig 2. Hence, the graphical representation below confirms and supports the results regarding the symposium.

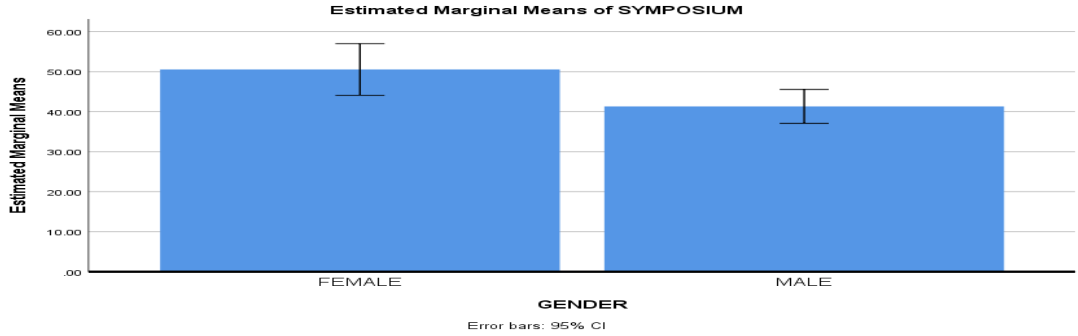

**Figure 2: Estimated Marginal Means of Symposium**

### 3.4.3 Applied Deployment

The significance level for applied deployment is 0.002 (Table 5), which is below the 0.05 level. This indicates a significant difference in the learning outcomes between female and male engineering students. The estimated margin levels are found to be 53 among females and 41 among males as shown in Fig 3. Hence, the graphical representation below confirms and supports the results regarding applied development.

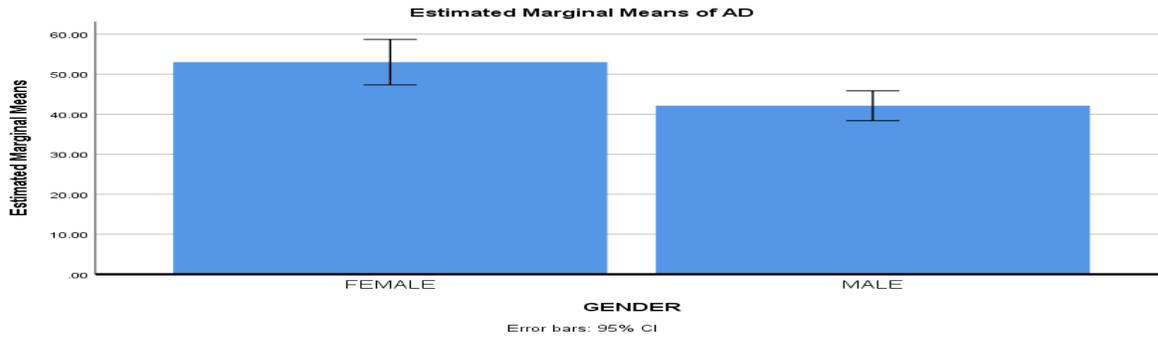

**Figure 3: Estimated Marginal Means of Applied Deployment**

### 3.4.4 Collaborative Learning

The significance level for collaborative learning is 0.106 (Table 5), which is below the 0.05 level. This indicates there is no significant difference in the learning outcomes between female and male engineering students. The estimated margin levels are found to be around 79 among females and 75 among males as shown in Fig 4. Hence, the graphical representation in Fig 4 confirms and supports the results regarding collaborative learning.

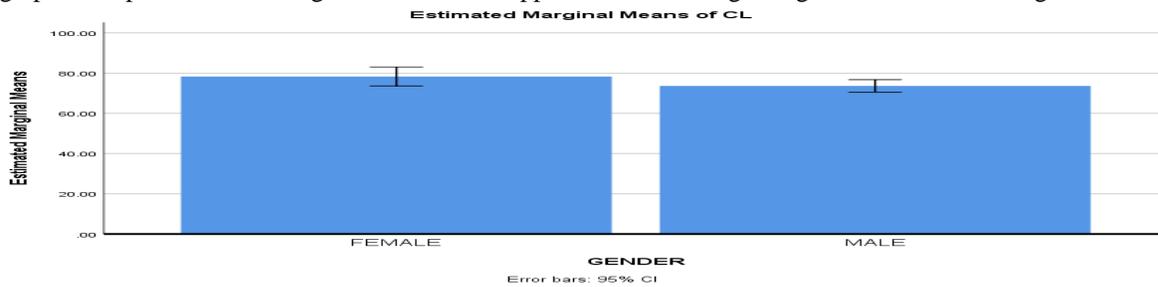

**Figure 4: Estimated Marginal Means of Collaborative Learning**

**3.5 Discussion**

The findings of this research are based on statistical analysis. As per the descriptive statistics, in terms of the assessment methods of direct conceptual learning, symposium, applied deployment, and collaborative learning, the average scores of female engineering students surpass those of their male counterparts. The multivariate tests further supported the findings of descriptive statistics indicating that the null hypothesis should be rejected. Cruz [14] supported the finding since the study explained that engineering students are supposed to attain an ability to work as competent team members in an interdisciplinary collaborative environment. It supports the need for implementing a collaborative learning environment in academics.

Alsalhi et al. [10] further supported the results since the authors stressed the importance of blended learning which further requires applied development learning. However, most studies in the literature have not emphasized the gender-based learning experiences of engineering students specifically. Hence, it indicates that the present research addressed a big gap in the literature. The present research further provided a unique perspective that learning outcomes can vary among different genders even if the same learning technique is applied.

The literature further stressed the importance of design practice [17]. It enables the students to implement the theoretical knowledge in real-world scenarios and devise core problem-solving skills. Similarly, the primary findings also support the importance of applied development among engineering students. The estimated margin levels indicated some difference among male and female engineering students indicating that both genders gained benefit from the applied development assessment methods. Ansari and Khan [30] also discussed the significance of collaborative learning among university students. It supports that collaborative learning can yield beneficial outcomes among both male and female students.

The critical analysis revealed that the literature has discussed the importance of learning assessment methods generally. There is a lack of evidence supporting the gender-based learning experiences of engineering students. It points out a gap in the literature that has been filled by the current research. The present research has also contributed to the academic discipline by presenting strong arguments regarding direct conceptual learning, applied development learning, collaborative learning, and symposiums. Further, future research can focus on at least 2-3 more learning assessment methods apart from those evaluated in this study.

However, the present research also comes with certain limitations like dependency on the primary quantitative method only. The absence of qualitative analysis and the limited scope of the research may have affected the results. Further and in-depth studies may be needed to confirm the findings. Also, there is a need to employ a large sample size in future research to develop generalized findings.

**4. CONCLUSION**

Teaching and learning are interconnected activities, and their effectiveness is influenced by various factors, including teaching methods, assessment practices, and their impact on student learning. The ultimate goal of teaching is to facilitate effective learning in students. The interplay between teaching methods and assessment practices can significantly influence student learning outcomes. When these components are thoughtfully designed and aligned, they enhance the learning process and promote deeper understanding, critical thinking, problem-solving skills, and overall engagement.

Our study provides conclusive evidence that there is a significant difference in the learning outcomes between male and female engineering students in the assessment methods of direct conceptual learning, applied deployment and symposium. In collaborative learning, the learning outcome is the same for both male and female engineering students. The statistical outcomes provided strong arguments that female engineering students outperform male students when using the mentioned assessment methods. It is also concluded that this research has addressed a literature gap. There is not enough evidence in the literature suggesting the core assessment methods for evaluating learning experience specifically among male and female engineering students.

In conclusion, the complex interplay between teaching methods, assessment practices, and their influence on student learning is critical for creating effective and efficient educational experiences. However, traits like dedication, zeal to learn, and constant improvise on the part of the student community add value to the efforts made by teachers in terms of upgrading teaching methods and skillfully designing assessment methods. These help engineering students keep up with the current market trends to prepare them better for their professional arena. These also play a role in gender-based learning experiences and enhancing the skill set. By carefully considering and optimizing these factors, educators can promote meaningful learning and achieve positive learning outcomes.